\magnification=\magstep1
 \hfuzz 6pc
 \def \e {{\rm e}}
 
\centerline{\bf PARTON FLUXES AND VIRTUAL PIONS}\vskip .5pc
\centerline{\bf  IN HEAVY NUCLEI}

\vskip 2pc
 \centerline {Giorgio Calucci 
 \footnote *{E-mail: {\tt giorgio@ts.infn.it} } }
 
 \vskip 2pc
\centerline{{\it Dipartimento di Fisica teorica dell'Universit\`a di Trieste,
 I 34014}}

\centerline{{\it INFN, Sezione di Trieste, Italy}}
\vskip 5pc

\centerline{\bf {Abstract} } \vskip 2pc
\narrower{ The partonic flux originated from a heavy nucleus is not the mere sum of the fluxes coming from the individual nucleons. There are various effects that give rise to modifications. One of these effects is here investigated $i.e.$ the presence of a cloud of virtual pions co-moving with the nucleus. It is found that the contribution of these virtual particles to the total parton flux should be rather small even if one includes the contribution of resonances.}

\vskip 1pc
 ${PACS\, numbers:\; 25.75.Ag\; 12.39.-x,}$
\vfill\eject

 \vskip 1pc
 {\bf 1. Introduction}\vskip 1pc
 In collisions at very high energy involving nuclei, in particular heavy
 nuclei, there is an interplay between high energy phenomenology which is
 described usually starting from different versions of parton models and the
 nuclear dynamics: the presence of many sources for the parton flux may induce relevant effects because they cannot be considered independent. These effects are usually studied directly at the partonic level and have therefore, at least in principle, their theoretical counterpart in QCD, even though it must be supplemented with phenomenological inputs to be able to produce answers at non perturbative level, see $e.g.$ [1]. Some of these investigations are moreover specifically devoted to heavy nuclei structure [2].\par
It is, however, clear that the nuclear interactions are not the direct expression of QCD, rather they are described by an intermediate dynamics which arises from the presence of a pion field interacting with the nucleons, so that the difference of the parton flux generated by a nucleus of number $A$, from the flux produced by $A$ independent nucleons arises also from the presence of interacting pions and, obviously, from the parton structure of the pions themselves.
\par
The aim of this note is to analyze this particular aspect of the parton structure of heavy nuclei by means of a two level model: a nuclear level in which the components are hadrons and a deeper partonic level in which the components are quarks and gluons. The
 model is not intended to reproduce details of the nuclear dynamics, but only
to take into account the fundamental aspects of the pion-nucleon interaction as
sources of the parton-flux modification. In a simplified dynamics, where there is only a quantized field of pions [3] linearly coupled to a source given by the 
nucleon, the effect is found to be very small. This result is easily recognized
to be due to the particular pseudoscalar coupling, so other forms of coupling
need to be investigated and this will be done by introducing the lightest of the
mesonic resonances. These dynamical models are then applied to the system of
many nucleons in order to see how the presence of many sources modify the meson population with respect to the simple sum of the populations pertaining to the individual nucleons. There could be a problem of partial double counting of the same degree of freedom, but what is attempted to calculate is not the total pionic population but only that part of it which differ from the mere sum of the pionic populations produced independently by the single nucleons. The problem of the pion population of the nucleus, as it is seen from scattering processes, has been already considered, see e.g. [4].
In order to be definite a precise dynamical model has been used, the one proposed in the book of J.D.Walecka [5].\par 
This is done in the next section and the overall effect is found to be small. For this reason the detailed partonic structure of the pion, which has been investigated since long time[6], has not been used in detail. The effect looked for could have some relevance in situation where many partonic interactions are relevant like in the so-called semi hard processes in nucleon-nucleus or nucleus-nucleus interactions, where the total sub-energy of the single collision is high enough to allow a perturbative treatment, but the fractional momentum is small, so the partonic population involved is large and the rescattering probability is large. Due to the roughness of the model no evolution with $Q^2$ in the parton distribution. is calculated. At the end a short estimate of the effects of Fermi motion and of  correlations is presented, and within the limits of the model it is possible to state that both effects must be really small for heavy nuclei.
 \vskip 1pc
{\bf 2. Virtual pions}\vskip 1pc
{\it 2.1 Overview of the single-nucleon models}\vskip 1pc
The models of Lewis, Oppenheimer and Wouthuysen [2] and of Bloch and Nordsieck [7] are well known but a short overview of them is needed for the future
discussions, in order to fix the notation and also because some modifications are introduced. (For a sketch of a different treatment, see Appendix). Both foresee a heavy spinor (the nucleon), which moves relativistically and interacts with lighter bosons. The nucleons cannot be created nor destroyed whereas the bosons are emitted and absorbed, so the nucleon is treated in first quantization, the bosons by means of a second quantized field. The interaction has the standard form
\footnote*{The arrow $\vec a$ is used to denote isovectors, the corresponding components take an index $u$, the boldface $\bf a$ denotes spatial three-vectors, and  $j$ is the corresponding index; the nucleons will bear the indices $i,\ell$, the different mesons the index $n$.}
 $${\cal L}_{\pi}=ig_{\pi}\bar \Psi \gamma_5 \vec\tau\cdot \vec\phi \Psi$$ for direct pion emission.\par As anticipated further interactions are considered: this will be done, following the suggestion by Walecka, by introducing resonances, the $\rho$ and the $\omega$ spin-one mesons and moreover a scalar-isoscalar $\sigma$ or $f_o$.
 So we have three further interaction terms:[5], respectively for the $\rho$, the $\omega$ and the scalar
 $${\cal L}_{\rho}=g_{\rho}\bar \Psi \gamma_{\mu} \vec\tau\cdot \vec U^{\mu}\Psi\quad \partial_{\mu} \vec U^{\mu}=0;\quad
 {\cal L}_{\omega}=g_{\omega}\bar \Psi \gamma_{\mu} \tilde U^{\mu}\Psi\quad \partial_{\mu} \tilde U^{\mu}=0\quad {\cal L}_s=g_s\bar \Psi \chi \Psi\;.$$
  The dynamics is limited in two ways: There in no emission or absorption of nucleons, the emitted particles carry a small amount of momentum, so that the velocity of the nucleon is not perturbed in a sizable way. All these limitations fit well with the aim of describing the emission and absorption of pions inside the nucleus. So the total Hamiltonian is:
$$ \eqalign{&{\cal H}_T= -i \alpha_j \partial_j+\beta M+\sum_n{\cal H}_{o,n}\cr & -ig_{\pi}\beta\gamma_5 \vec\tau\cdot \vec\phi-g_s \beta\chi-g_{\rho}\vec\tau\cdot[ \alpha_j\vec U_j-\vec U_o]-g_{\omega}[ \alpha_j\tilde U_j-\tilde U_o]\,, }\eqno (1)$$ 
where ${\cal H}_{o,n}$ are the free Hamiltonian of the pions, of the scalars, 
 of the $\rho-$mesons and of the $\omega-$mesons, $i.e.\;n=\pi,s,\rho,\omega.$
  For the masses and the energies the notations are: $M,\;E$ for the nucleon, $m_n\;\epsilon_n$ for the mesons. The wave function of the nucleon is written as $\Psi({\bf r})=\psi \e^{i{\bf p\cdot r}}$, where:
  $$\psi=\sqrt{{E+M}\over {2E}}\left\|
  \matrix{\hfill\xi\hfill\cr
   {{\bf\sigma\cdot p}\over{E+M}}\xi\hfill   }\right\|.$$
   The normalization $\xi^*\xi=1$ gives $\psi^*\psi=1$.
   The Dirac matrices are evaluated between plane waves of given velocity
   with the known results:
   $$<\alpha_j>=v_j\quad <\beta>=1/\gamma \quad <\beta\gamma_5>=\sigma\cdot {\bf  \hat k} /M\gamma\;. \eqno (2)$$
   The first non zero result for $<\beta\gamma_5>$ is obtained at the first order in the momentum transfer; ${\bf \hat k}\equiv (k_{\perp},k_{\|}/\gamma)$ and the directions transverse and parallel are defined with respect to the velocity $\bf v$. The vector fields are decomposed taking into account the subsidiary condition $\partial_{\mu} U^{\mu}=0$, so that only three polarizations enter[8]. At this point there are still matrix structures, the spin matrices $\sigma_j$ and the isospin matrices $\tau_u$, they make the subsequent calculation much more complicated than in Q.E.D. [7].
They will be substituted by numerical vectors $s_j\,,t_u$ corresponding again to some mean value over the nucleon state; the normalization is such that $s_j$ is twice the spin and $t_u$ twice the isospin of a nucleon.
 In so doing the original Hamiltonian becomes:
  $$ {\cal H}_T= -i {\bf v\cdot \partial}+M/\gamma+\sum_n\int\!\Big\{\epsilon_{n,k} a_n^{\dagger}({\bf k}) a_n({\bf k}) -\big[ a_n({\bf k}) J_n({\bf k})+a_n^{\dagger}({\bf k}) J^*_n({\bf k })\big]\Big\}d^3k \eqno(3)$$
The sources $J_n$ have the explicit form:$$\eqalign{
J_{\pi}=&{ig_{\pi}\over {(2\pi)^{3/2}}}{1\over{\sqrt{2\epsilon}}}{1\over{M\gamma}}{\bf s\cdot\hat k}\;\vec t\e^{i{\bf k\cdot r}}\cr
J_{s}=&{g_s\over {(2\pi)^{3/2}}}{1\over{\sqrt{2\epsilon}}}{1\over{\gamma}}\e^{i{\bf k\cdot r}}\cr
J_{\rho}=&{g_{\rho}\over {(2\pi)^{3/2}}}{1\over{\sqrt{2\epsilon}}}V_j\vec t\,\e^{i{\bf k\cdot r}}\cr
J_{\omega}=&{g_{\omega}\over {(2\pi)^{3/2}}}{1\over{\sqrt{2\epsilon}}}V_j\e^{i{\bf k\cdot r}}\cr} \eqno(4)$$
The vectorial coefficients $V$ are given by
$$ V_j={\bf e_j\cdot v},\; {\bf k\cdot e_j}=0,\; {\rm for\;} j=1,2\eqno(5a)$$
$$ V_3={{\epsilon\, {\bf k\cdot v}}\over {m|k|}}-{{|k|}\over m}\eqno(5b)$$
It is known that the total Hamiltonian is brought to diagonal form by means of the transformation:
$$ {\cal U} a_n {\cal U}^{\dagger}=b_n \quad {\rm where}\quad {\cal U}=\exp\sum_n\int [a_n F_n-a_n^{\dagger} F^*_n]d^3k$$
with:
$$ F_n=f_n\,\e^{i{\bf k\cdot r}}={{J_n}\over{\epsilon_n-{\bf v\cdot k}_n}}$$
In this way, in the lowest level of ${\cal H}_T$, characterized by $b_n({\bf k})|\circ>=0$, the meson population is given by
 $$<\circ|a_n^{\dagger}a_n|\circ>=|F_n|^2=|f_n|^2\eqno (6)$$
 The total population is then obtained by summing over the internal (spin and isospin) quantum numbers.\par
  The populations are independent of $\bf r$, they depend, however, on $\bf v$.

 \vskip 1pc
{\it 2.2 Many nucleons}\vskip 1pc
 In a heavy nucleus, with $A$ nucleons, there are many nucleons with negligible relative
 velocities, equal masses, but located at different points, the new Hamiltonian is easily obtained from eq (2):
  $$\eqalign{ {\cal H}_A= &\sum_{\ell=1}^A[-i {\bf v\cdot \partial}_{\ell}+M/\gamma]+\sum_n\int \epsilon_{n,k} a_n^{\dagger}({\bf k}) a_n({\bf k}) d^3k\cr
 -&\sum_{\ell,n}\int\big[ a_n({\bf k})J_n({\bf k,r_{\ell}})+a_n^{\dagger}({\bf k}) J^*_n({\bf k,r_{\ell} })\big]d^3k }\eqno(7)$$ 
  The procedure needed in order to make diagonal the new Hamiltonian is the same as before, with a new operator ${\cal U}_A$ and
 $${\cal U}_A= \exp\sum_{\ell,n}\int [a_n F_{\ell,n}-a_n^{\dagger} F^*_{\ell,n}]d^3k.$$ 
The population of the mesons of kind $n$ is now given (see previous eq. (6)) by: 
$$<\circ|a_n^{\dagger}a_n|\circ>=\Big|\sum_{\ell} F_{\ell,n}\Big|^2=
\sum_{\ell}|f_{\ell,n}|^2+\sum_{\ell\neq i} f^*_{\ell,n}f_{i,n}\e^{i{(\bf r_i-r_{\ell})\cdot k}} \eqno (8)$$
 The two addenda have quite different meanings, the first one is just the description of the meson clouds of the independent nucleons, they are simply part of the physical nucleons,(see $e.g. [9]$) the second term says how the presence of many nucleons modifies the meson population, so it describes the effect that is looked for, it will be indicated in general as ${\cal N}_n$ and studied more in detail: The positions $\bf r$ are defined in the frame where the nucleon is rapidly moving, see eq. (3), so the nucleus appears as a
sphere contracted in the direction of the motion. These positions must be mediated over the nuclear volume, to this end it is convenient to define
${\bf \check r }\equiv (r_{\perp}, r_{\|}\gamma )$ so that with respect to  ${\bf \check r}$
the volume is spherical, moreover ${\bf\check r\cdot\hat k=r\cdot k }$. The exponential factor, mediated over a sphere of radius $R$ gives the coherence factor
 $$C(\hat k R)=\int_{\check r\le R}\e^{i{\bf \hat k\cdot\check r}} d^3\check r\bigg/\int_{\check r\le R} d^3\check r={3\over {(\hat k R)^2}}\Big[
 {{\sin(\hat k R)} \over{\hat k R}}-\cos(\hat k R)\Big]\,.\eqno(9) $$
The two oscillating factors must be mediated separately and therefore the factor ${\cal M}=C^2$ appears finally. Note that $R\simeq A^{1/3}\,r_o$, with $r_o$ radius of the nucleon.\par
There are other factors that must be averaged over the whole nucleus: the isospin factor in the case of single-pion emission and of the $\rho-$resonance and also the spin factor in the case of single-pion emission.
For the isospin one can start from [9]:
$$(T_u)^2=\Big(\sum_i t_{i,u}\Big)^2= \sum_i (t_{i,u})^2+\sum_{i\neq \ell}t_{i,u}t_{\ell,u}$$
and one must estimate the mean value of $(T_u)^2$. By definition
$T_3= Z-N<0$ so $(T_3)^2= (Z-N)^2$; the mean value of $(T_u)^2\; u=1,2$ is not so easily defined, in every case:
$$\big<\vec t_i\cdot \vec t_{\ell}\big>\geq {{ (Z-N)^2- 3A}\over {A^2}}. \eqno
(10)$$ 
 The spin of the nucleon is treated in an analogous way: fixed an
arbitrary direction $j$ in the space $ \hat k$ one can write;
$\sum_i s_{i,j}=S_j$, defining the component $j$ of twice the total spin of the nucleus. This value is certainly very small, it does not grow with $A$, so the result is
$$(S_j)^2= \Big(\sum_i s_{i,j}\Big)^2= \sum_i (s_{i,j})^2+\sum_{i\neq \ell}s_{i,j}s_{\ell,j}$$
 As a mean value  $<(s_{i,j})^2>=1$ and $<(S_j)^2>$ is of the same order 1, so that, at the end $$\big<s_{i,j}s_{\ell,j}\big>\approx -1/A. \eqno (11)$$
 
 In order to write the 
 expression for ${\cal N}_n$ for the meson populations for a given value 
 of the momentum $\bf k$, we must still sum over the polarizations of the vector
  particles: $\sum \big|V_j\big|^2=(\epsilon-{\bf v\cdot k})^2/m^2-1/\gamma^2$.  The sum $\sum_{\ell\neq i}$ gives simply a factor $A(A-1)\simeq A^2$, so the final form is:$$
    {\cal N}_{\pi}=-{{g_{\pi}^2}\over {(2\pi)^3}} {{(N-Z)^2-3A}\over A}
    \Big({{\hat k}\over M}\Big)^2 {1\over{\gamma^2 (\epsilon-{\bf v\cdot k})^2}} {1\over {2\epsilon}}C^2(\hat k R)\eqno (12 a)$$
   $$ {\cal N}_s={{g_s^2}\over {(2\pi)^3}}  A^2     {1\over{\gamma^2 (\epsilon-{\bf v\cdot k})^2}} {1\over {2\epsilon}}C^2(\hat k R)\eqno (12 b)$$
 $$ {\cal N}_{\rho}={{g_{\rho}^2}\over {(2\pi)^3}} \big[(N-Z)^2-3A\big]   \bigg[{1\over m^2}-{1\over{\gamma^2(\epsilon-{\bf v\cdot k})^2}}\bigg]{1\over {2\epsilon}} C^2(\hat k R)\eqno (12 c)$$
 $$ {\cal N}_{\omega}={{g_{\omega}^2}\over {(2\pi)^3}} A^2   \bigg[{1\over m^2}-{1\over{\gamma^2(\epsilon-{\bf v\cdot k})^2}}\bigg]{1\over {2\epsilon}} C^2(\hat k R) \eqno (12 d)$$
    Due to the rapid fall off at large $k$ of the factor $C$ it is possible to perform the complete integration over the momenta of the mesons obtaining the total population. \par
For the direct production of pions (eq.12 a) the factors coming from spin and isospin give rise to an overall coefficient smaller than in the other three cases. The emission probability, however, increases with decreasing mass, so also the single pion production is considered.
  
   \vskip 1pc
{\it 2.3 Total population of pions}\vskip 1pc

The total pion number can be estimated by computing ${\cal T}_n=\int {\cal N}_n d^3k$, in detail:
  $$\eqalign{{\cal T}_{\pi}=&-{{g_{\pi}^2}\over {(2\pi)^3}} \big[(N-Z)^2/A-3\big]\cdot\Big({{m_{\pi}}\over M}\Big)^2{\cal W}_o(m_{\pi})\cr
{\cal T}_s=&{{g_s^2}\over {(2\pi)^3}} A^2 \cdot{\cal W}_1(m_s)\cr
{\cal T}_{\rho}=&{{g_{\rho}^2}\over {(2\pi)^3}} \big[(N-Z)^2-3A\big]\cdot\big[{\cal W}_2(m_{\rho})-{\cal W}_1(m_{\rho})\big]\cr
{\cal T}_{\omega}=&{{g_{\omega}^2}\over {(2\pi)^3}} A^2 \cdot\big[{\cal W}_2(m_{\omega})-{\cal W}_1(m_{\omega})\big] \cr }  \eqno (13)$$
 where:
 $$
    {\cal W}_o(m)=\int\Big({{\hat k}\over {m\gamma}}\Big)^2 {1\over{2\epsilon (\epsilon-{\bf v\cdot k})^2}} C^2(\hat k R)\,d^3k \eqno (14 a)$$
   $$ {\cal W}_1(m)=\int\Big({1\over {\gamma}}\Big)^2 {1\over{2\epsilon (\epsilon-{\bf v\cdot k})^2}} C^2(\hat k R)\,d^3k 
   \eqno (14 b)$$
 $$ {\cal W}_2(m)=\int{1\over m^2}{1\over {2\epsilon}} C^2(\hat k R)\,d^3k)\;.\eqno (14 c)$$
 There are two well separated contribution to the meson population. 
In ${\cal W}_o$ and ${\cal W}_1$ there is a rapidly varying denominator which finds its minimum in configurations satisfying the relation: ${{\partial}\over {\partial k_{\|}}} [\epsilon -vk_{\|}]=0\; i.e.\;k_{\|}=v\epsilon $, where the following relations hold:
$$ \epsilon=\gamma m_{\perp},\quad k_{\|}=v \gamma m_{\perp},\quad
\hat k^2=v^2 m_{\perp}^2+k_{\perp}^2,\quad {\rm with}\quad m_{\perp}^2=k_{\perp}^2+ m^2\,. \eqno(15)$$
Since the denominator varies rapidly the whole integral is estimated by using the standard mean-value approximation: the value $k_{\|}=v\epsilon$ is taken in all the factors of the integrand, except in the rapidly varying denominator which is integrated exactly in $k_{\|}$. 
$$ \int {{ dk_{\|}}\over {2\epsilon (\epsilon -vk_{\|})^2}}=
{{\gamma^2}\over{ m_{\perp}^2}}{{1+v}\over 2}\approx {{\gamma^2}\over{ m_{\perp}^2}}\,.$$
Moreover, from eq.s (14) it follows that $d k^2_{\perp}/m^2_{\perp}=d\hat k^2/[\hat k^2+m^2]$ and the lower limit $k^2_{\perp}=0$ correspond to the lower limit $ \hat k^2=(mv)^2$. In conclusion we get the following expression for the integral (with $t=\hat k/m$): 
$$ {\cal W}_o=2\pi \int_{v}^{+\infty} C^2(mRt) {t^3\over {t^2+1}} dt\quad {\cal W}_1=2\pi \int_{v}^{+\infty} C^2(mRt) {t\over {t^2+1}} dt \;.
\eqno (16) $$
The population described by $ {\cal W}_1$ and $ {\cal W}_o$ is therefore composed by mesons co-moving with the nucleons of the nucleus, so it is precisely the effect that was looked for and which will modify the partonic population.\par
In ${\cal W}_2(m)$, on the contrary, there is no rapidly varying denominator, the expression  is symmetrical between forward and backward direction: when seen from the point of view of $\bf r-$coordinates it corresponds to a mesonic field, whose source becomes more and more coherent when the whole nucleus is squeezed, so its subsequent interpretation in terms of partons is doubtful, anyhow the result can be calculated and takes a simple form in the limit $\gamma\to\infty$.
By carrying out the integration over the solid angle (of the variable $\hat k$)  the result is
$${\cal W}_2={{4\pi}\over{(mR)^2 v}}\int_0^{\infty}  C^2(x)\arcsin\!{\rm h} \Big[ {{xv\gamma}\over{\sqrt{x^2+(m R)^2}}}\Big]xdx\quad {\rm
(here} :\;x=\hat kR) \eqno(17)$$
and the limit of large $\gamma$ gives
$${\cal W}'_2={9\over 4}\ln 2\,\gamma {{4\pi}\over{(mR)^2 }}\int_0^{\infty}  C^2(x)x dx=\ln 2\gamma {{4\pi}\over{(mR)^2 }}\,. \eqno(18)$$
The present investigation is intended only to give indications about the order of magnitude of the effect, anyhow, some numerical estimates of the values of the pion population are presented. \par
The total number of co-moving pions is given as:
$$ {\cal P}={\cal T}_{\pi}+2{\cal T}_s+2{\cal T}^{(1)}_{\rho}+3{\cal T}^{(1)}_{\omega} \;.\eqno (19)$$
The apex ${}^{(1)}$ indicates that only the contribution of ${\cal W}_1$ is taken.\par
 As typical heavy nuclei the gold and the lead are chosen, they have respectively $A=197,\;N-Z=39$ and $A=208,\;N-Z=44$.
 The masses and the coupling constants are taken from ref [5] as:
 $m= 138,\;=550,\;=782$MeV for the pion, for the scalar and for the 
 $\rho$ and $\omega$ and
 $g^2/(4\pi)= 14.4,\;=7.3,\;=10.8$ for the pion, for the scalar and for the 
 $\rho$ and $\omega$; the mass and the radius of the nucleon are 939MeV and 1.07fm, so that $(m_{\pi}/M)^2=0.022$.
 Some numerical estimates of the values of ${\cal W}_o$ and ${\cal W}_1$  for the relevant values of the parameters are needed and then the final result is obtained in the form 
 $$ {\cal P}\Big|_{Au}=0.58\quad{\cal P}\Big|_{Pb}=0.73\;. \eqno (20)$$
 These numbers are small not only because of the smallness of the single addenda but also because not all the addenda have the same signs, see eq (13). The ratio of the number of co-moving pions with the number of nucleons is in both cases around $0.3\%$, which puts in better evidence the smallness of the effect.\par
 The terms ${\cal W}_2$ give a larger contribution, but the corresponding mesons are not moving together with the nucleus, they cannot contribute to the usual parton population.
\vskip 1pc
  {\bf  3. Other effects and corrections}
\vskip 1pc
 {\it  3.1 Fermi motion}\vskip 1pc
 The Fermi motion of the nucleons inside the nucleus gives rise to a variation on the partonic variables when seen in the nuclear-c.m.frame. In particular the (+)variables are multiplied by a factor depending on the rapidity of the parent nucleon. This factor is $\e^{\eta}$ and the rapidity $\eta$ in the actual case takes small values, of the order of $p_F/M$. So if a nucleon at rest with respect to the nucleus yields a distribution $w(x)$, taking into account the relative motion the distribution becomes $w(x\e^{\eta})$. It is useful to expand the expression in $\eta$ and to perform an average over the different longitudinal relative motion, which are evidently symmetric about the $\eta=0$ case. For the low$-x$ part of the partonic spectrum, which should behave as $x^{-\nu}$ the modification (after averaging) takes the form of a multiplication by $1+\nu^2 p_F^2/6M^2$ and is therefore quite small.\par
The same argument applies to the pion population: the maximum of the function $xC^2(x)$ is located around $x=x_o\approx 2.5$, this implies that the maximum of the population is found for $\hat k^2=k_{\perp}^2+k_{\|}/\gamma^2=x_o^2/R^2$, so here also $k_{\perp}$ shall be small and $k_{\|}$ grows with $\gamma$ so it is (mainly) co-moving with the nucleus. The argument should be refined by considering that the large majority of the pions comes from the resonances so there is further boost and dispersion in their spectrum. 
\vskip 1pc
 {\it  3.2 Correlations}\vskip 1pc
The presence of correlations among the nucleons can influence the partonic structure of the nucleus. In order to deal with the correlations one can start with a general form of the nucleon distribution[10].
$$\rho({\bf r_1,r_2,\dots r_n})=\prod_i \rho^{(1)} ({\bf r_i})+\sum_{j\neq k}\Delta({\bf r_j,r_k})\prod_{ i\neq j,k} \rho^{(1)}({\bf r_i})+\dots $$
with the condition $\int d^3r_k \Delta({\bf r_j,r_k})=0$. Then taking for $\rho^{(1)}({\bf r_i})$ a constant and the correlation radius in $\Delta$ much smaller than the nuclear radius one gets, instead of ${\cal M}=C^2$ eq(9), the more complicated expression;
$${\cal M}'=C^2(\hat k R)+{3\over{4\pi R^3}}\int d^3 u \Delta(u) \e^{i\hat k\cdot u}\,.$$ 
 The difference of ${\cal M}'$ from ${\cal M}$ depends on the particular form of the correlation, but until the correlation range $a$ is sizably smaller than $R$ the corrective term is of the order $(a/R)^3$, there is no reason for $a$ to grow with $A$, the corrections decrease with $1/A$.\par
Another kind of correlation could be present, the one between spin and isospin. In eq.s(10,11) the spin and isospin were averaged independently. Now one remembers that the two-nucleon wave function must be totally antisymmetric. Usually it happens that the symmetric wave function in the space coordinates corresponds to a lowest value of the energy, so it is favored. In this case the antisymmetry is shifted to the spin-isospin term, inducing therefore a correlation. However this effect can be relevant only for the single-pion emission which gives the minor contribution to the whole effect.\par 
Finally one could inquire which could be the effect of other many-pion states. The obvious next candidate would be the $\eta-$resonance, decaying into three pions. Since it is pseudoscalar and isoscalar its contribution would have the form:  $${\cal T}_{\eta}=-{{g_{\eta}^2}\over {(2\pi)^3}}A \cdot\Big({{m_{\eta}}\over M}\Big)^2{\cal W}_o(m_{\eta}).$$ 
With a mass of 547MeV and a coupling $g_{\eta}^2/(4\pi)$ of the order of ten one gets a contribution around $10^{-2}$ to the total population $\cal P$, so the possible correction is really tiny. The effect of higher resonances is certainly even smaller, note moreover that possible contributions from axial resonances are depressed by the spin averaging process.
\vskip 1pc
 {\it  3.3 Conservation of the vector current}\vskip 1pc
In some phenomenological treatments one considers the vector meson coupled to {\it conserved} vector currents $e.g.$ the $\rho$-meson to the isospin current. Here a brief investigation is performed on this version because in so doing one gets also some insight on the possible meaning of the unpleasant term ${\cal W}_2$.\par
If one enforces the vector current to be conserved the coupling of the vector meson instead of being as in eq.(1) $g[\alpha_j U_j-U_o]$ will be, in momentum space, $$g[\alpha_j U_j-\alpha\cdot {\bf k}U_o/\epsilon] \eqno(21)$$  
As a consequence the vector coefficient $V_3$ given in eq.(5b) will be changed to
$$ V_3={{\epsilon\, {\bf k\cdot v}}\over {m|k|}}-{{|k|}\over m} {{{\bf k\cdot v}}\over {\epsilon}}={{m{\bf k\cdot v}}\over {\epsilon|k|}}\eqno(5c)$$
Hence the sum of the squares is $D=\sum \big | V_j\big |^2=v^2-({\bf v\cdot k})^2/\epsilon^2$.
The number of mesons generated by a conserved current, (with some coefficient $Q$ depending on $N$ and $Z$) can be put in the following form:

 $$ {\cal N}_c={{g^2}\over {(2\pi)^3}} Q\bigg[{1\over \epsilon^2}{{\epsilon+{\bf v\cdot k}}\over{{\epsilon-{\bf v\cdot k}}}}-{1\over{\gamma^2(\epsilon-{\bf v\cdot k})^2}}\bigg]{1\over {2\epsilon}} C^2(\hat k R)\;; $$
thus every term has a non constant denominator, but the in the new addendum the variation is slower. If we evaluate the sum in the minimum of the denominator it is easily found $D=1/\gamma^2$, so the result is the same as given before by the term ${\cal W}_1$, but {\it with opposite sign} and no contribution of the type ${\cal W}_2$ is found.\par
 In conclusion we can relate the term ${\cal W}_2$ to the well known bad behavior of a vector field coupled to a non-conserved current. From a  phenomenological point of view there is no obvious reason for requiring the source of massive vector mesons to be conserved, two perhaps lucky facts happen in this particular case: the term ${\cal W}_2$ gives a contribution which is kinematically well separated from the other ones; the request that the current be conserved yields a result which is different in sign from the previous one,
so that the values of $\cal P$ given in eq.(20) are now larger, around 2, but the property that the overall effect is small is well preserved.

 \vskip 2pc

 {\bf 4. Conclusions}\vskip 1pc
An aspect of the parton flux for relativistic heavy nuclei has been examined: it has to do with the presence in the nucleus of virtual pions, which are boosted to relativistic speed together with the component nucleons. It is found that 
the overall relevance of the pions is small; the effect is comparable with the variation of one in the total number of the nucleons in the nucleus.
How the shape in $x$ of the parton flux is modified in comparison with a free-nucleon flux is a more complicated question since it depends on the details of the partonic structure function of the pion. It has already been noted that investigations on this problem are available[6] a particular interest being addressed to the distribution of the valence quark [11], whereas all the collective effects should become more important for small$-x$ partons.  
The effect of pion produced through resonances, or in pair, is more relevant than the effect of single-pion production, which, by itself, gives a negative effect as a consequence of the spin structure of the interaction.\par
There is also an effect due to the superposition of the parton fluxes, the terms previously called ${\cal W}_2$, but it has not the correct kinematical properties. The superposition of the fluxes due to individual nucleons and the partial fusion of them can be studied [1] even without taking care of the contributions coming from the virtual pions: they are there, but  their contribution never exceeds some per cent to the final total flux. It is pleasant that the qualitative final result: the total effect is small, is stable with respect to various model dependences and incertitudes.

 \vskip 2pc
 {\bf Appendix}\vskip 1pc
 In this paper an old-fashioned formalism is employed, so it could
 be useful to outline a procedure that makes use of the more common language of
 the Feynman graphs, in particular in connection with the "infinite momentum"
 frame [12]. Since the main issue is the examination of the interference terms
 in the meson production from different nucleons it is enough to consider two
 interacting nucleons. The spin effects are not considered here, only the
 relevant singularity structure is examined.
 This different procedure puts in a clearer way the kinematical approximations that were more implicit in the previous treatment.\par
 We may start from a two-body (amputated) Bethe-Salpeter function:
 $ \Phi(p_1,p_2)$; defining 
 $k=p_1-p_2,\;P=p_1+p_2+q $  we can also write $\phi_P(k)$. If we want to
 include the possibility of emission of a light particle, which here is for
 simplicity treated as a scalar one, we get two terms:
  $$\Phi(p_1+q,p_2)g {1\over{(p_1+q)^2-m^2}}\quad ,  \quad 
\Phi(p_1,p_2+q)g {1\over{(p_2+q)^2-m^2}}\,. $$ Now the total momentum is $P=p_1+p_2+q$, the previous expression can be rewritten as:
$$\phi_P(k+q) g {1\over{(P+k)\cdot q}}\quad , \quad \phi_P(k-q)g{1\over{(P-k)\cdot q}}\;.$$
The production probabilities are given by the absolute square of the above expressions, but as discussed previously the production by a single nucleon corresponds to the "dressing" of these particle with lighter mesons, the relevant term is the interference, which distinguishes the bound nucleon from the free one and this term is precisely:
$${\cal J}=g^2\int d^4k\;\phi_P(k+q) {1\over{(P+k)\cdot q}} {1\over{(P-k)\cdot q}} \phi^*_P(k-q)\;.\eqno (A.1)$$
 The integration variable $k$, which says how much the nucleons are off mass shell, is controlled by the wave function $\phi_P$ and cannot become too large, so one expands the product of propagators as:
 $$ {1\over{(P+k)\cdot q}} {1\over{(P-k)\cdot q}}\approx {1\over {(P\cdot q)^2}}+{{(k\cdot q)^2}\over{(P\cdot q)^4}}+\cdots \eqno(A.2)$$ and represents the wave functions as $$\eqalign{
  \phi_P(k+q)=&(2\pi)^{-2}\int f_P (x)\exp[ix\cdot(k+q)]d^4x\cr
 \phi^*_P(k-q)=&(2\pi)^{-2}\int f^*_P (y)\exp[-iy\cdot(k-q)]d^4y\,.} \eqno (A.3)$$

Calculating ${\cal J}$ by means of the representation (A.3) the first term in the expansion (A.2) ${\cal J}_o$, is:
  $${\cal J}_o= {{ g^2}\over{(P\cdot q)^2}}\int d^4x \big|f_P(x)\big|^2\e^{2ix\cdot q}\;.\eqno (A.4)$$
  The first correction is more complicated but can also be computed. In analogy with eq(A.3) it results
$$ k\cdot q\,\phi_P(k+q))=\int d^4x f_P(x)\e^{ix\cdot q}(-iq\cdot\partial_x)\e^{ix\cdot k} \eqno (A.5) $$ 
and similarly for $\phi^*$. The function $f_P$ goes certainly to zero together with its derivatives at infinity, so some partial integration may be performed.
In this way the first correction ${\cal J}_1$ is found to be:
$${\cal J}_1= {{g^2}\over {(P\cdot q)^4}}\int d^4 x
\Big[q^4  \big|f_P(x)\big|^2+ \big|q\cdot\partial_x f_P(x)\big|^2+2\Im\Big(q\cdot\partial_x f^*_P(x)q^2 f_P(x)\Big)\Big]\e^{2ix\cdot q}\; \eqno (A.6) $$
Now in order to recover the previous result one must specialize the function $f_P$; the particular form of eq.(9) is produced by reducing the four-dimensional dependence on $x$ to a three-dimensional one (no relative time[13]), reasonable because the internal nuclear dynamics is not extremely relativistic, the spatial shape was then expressed by a sharp-edge ball instead of a more realistic distribution just in order to simplify the estimate. The kinematical situation in eq.s (3) and (7) corresponds to $P_o,P_3 \to\infty$ at fixed $P_{\bot}$.\par
The effect of the correction (A.6) may be expressed as:
$${\cal J}_o+{\cal J}_1={\cal J}_o\times{{ q^4}\over{(P\cdot q)^2}}+{\cal J}_{\rm s}$$ It appear then clearly that the correction is larger for larger masses $\mu^2=q^2$ of the produced particles,
whereas ${\cal J}_{\rm s}$ describes to terms depending on the shape of the nuclear surface.
\vfill\eject

{\bf References}\vskip 1pc
1 - I.P.Ivanov, N.N.Nikolaev, W.Sch\"afer, B.G.Zacharov, V.R.Zoller \par {\it Lectures on Diffraction and Saturation of Nuclear Partons in DIS off Heavy Nuclei} arXiv:hep-ph/0212161v2 (2003)\par
    \quad A.S.Rinat, M.F.Taragin Phys. Rev. C72 065209 (2005)\par
    \quad J.Rozinek, G.Wilk Phys. Lett.B473, 167 (2000)\par
    \quad S.Kumano, K.Umekawa {\it Modification of parton distribution in nuclei} arXiv:hep-ph/9803359 (1998)\par
    \quad M. Hirai, S. Kumano, T.-H. Nagai Phys.Rev. C70 044905 (2004) and {\it Nuclear corrections of parton distribution functions} arXiv:hep-ph/0408135 (2004)
2 - R.Venugopalan Nucl.Phys A590, 147c (1995)\par 
3 - H.W.Lewis, J.R.Oppenheimer, S.A.Wouthuysen  Phys. Rev. 73, 127 (1948)\par
4 - G.E.Brown, M.Buballa, Zi Bang Li, J.Wambach Nucl.Phys A593, 295 (1995)\par
    \quad G.A.Miller {\it Low energy pion nucleus interaction: Nuclear deep inelastic scattering, Drell Yan and missing pions} arXiv:nucl-th/9611043 (1996)\par 
5 - J.D.Walecka  Theoretical nuclear and subnuclear Physics - Ch. 20 World Scientific 2004\par 
6 - G. Altarelli, N. Cabibbo, L. Majani, R. Petronzio Nucl. Phys. B92, 413 (1975)\par\quad
M. Lusignoli  and P. Pistilli Lett. N. Cimento 24, 81 (1979)\par\quad
M. Lusignoli, P. Pistilli, F. Rapuano Nucl. Phys. B155, 394 (1979)\par\quad
J. Speth  and A.W. Thomas Adv. Nucl.Phys.24, 83 (1997)\par\quad
 B.C.Tiburzi, G.A.Miller Phys.Rev D67 013010 {\it and} 113004 (2003)\par
7 - F.Bloch and A.Nordsieck Phys Rev.52, 54 (1937)\par
    \quad W.Thirring and B.Touschek Phil. Mag. Ser.7,Vol.XLII, 244 (1951)\par
8 - H.Umezawa  Quantum Field Theory  North Holland 1956\par
9 - R.L.Sugar Phys. Rev D9, 2474 (1974)\par 
10 - M.Alvioli, C.Ciofi degli Atti, I.Marchino, V.Palli, H.Morita Phys. Rev. C78, 031601R (2008)\par
11 - M.Alberg and E.M.Henley Phys.Lett.B 611, 111 (2005)\par
12 - S.Fubini  and G.Furlan  Physics 1,229 (1965)\par\quad
    S.Weinberg Phys. Rev. 150,1313 (1966)\par
13 - E.E.Salpeter Phys. Rev. 87,328(1952)
\vfill\eject\end